\documentclass[12pt,sort&compress]{article}

\usepackage[dvips]{color}
\usepackage{amsmath}
\usepackage{graphicx} 
\usepackage[dvips]{color}
\usepackage{graphicx}
\usepackage{tabularx}
\usepackage{subfig}
\usepackage{cite}
\usepackage{amsfonts}
\usepackage{tabularx,ragged2e}
\newcolumntype{C}{>{\Centering\arraybackslash}X} 
\usepackage[outercaption]{sidecap}    
\usepackage{hyperref}

\newcommand{\gammast}{{\gamma^\ast}}

\begin{document} 
\title{Susceptibility to disorder of the optimal resetting rate in the Larkin model of directed polymers}
\date{}
\author{Pascal Grange\\
Department of  Physics\\
 Xi'an Jiaotong-Liverpool University\\
111 Ren'ai Rd, 215123 Suzhou, China\\
\normalsize{{\ttfamily{pascal.grange@xjtlu.edu.cn}}}}
\maketitle


\begin{abstract}
We consider the Larkin model of a directed polymer with Gaussian-distributed random forces, with the addition of a resetting process whereby the transverse position of the end-point of the polymer is reset to zero with constant rate $r$. We express the average over disorder of the mean time to absorption by an absorbing target at a fixed value of the transverse position. Thanks to the independence properties of the distribution of the random forces, this expression is analogous to the mean time to absorption for a diffusive particle under resetting, which possesses a single minimum at an optimal value $r^\ast$ of the resetting rate . Moreover, the mean time to absorption can be expanded as a power series of the amplitude of the disorder, around the value $r^\ast$ of the resetting rate. We obtain the susceptibility of the optimal resetting rate to disorder in closed form, and find it to be positive. 
\end{abstract}

\tableofcontents

\section{Introduction}

  In a one-dimensional  search problem, one may cut off  long excursions into the wrong direction 
  by returning to the starting point after a duration 
 of search considered excessive. In a model of such a strategy, the searcher
 was assimilated  in \cite{evans2011diffusion} to a diffusive random walker on a line with an absorbing target, resetting its 
 position to its starting point at stochastic times, distributed according to a Poisson law of fixed rate $r$.  
  The corresponding non-equilibrium stationary state was worked out.  Morover, the mean first-passage time at the target
 is made finite by the resetting process, and can be minimised by adjusting the  resetting rate to a value 
 proportional to the diffusion constant divided by the square of the distance between the initial position and the target, up to a 
  numerical constant given in terms of a transcendental equation \cite{evans2011optimal}.\\
  
   Stochastic resetting has since become a source
 of developments  in out-of-equilibrium statistical physics \cite{majumdar2015dynamical,maes2017induced,masoliver2019anomalous,IsingResetting},
 with applications including RNA polymerisation processes \cite{roldan2016stochastic,lisica2016mechanisms}, active matter \cite{scacchi2018mean,evans2018run,masoliver2019telegraphic}, randomised searching problems \cite{kusmierz2014first} and lifting of
 entropy barriers \cite{grange2019entropy}. 
 The corresponding renewal arguments \cite{evans2011diffusion,evans2011optimal,evans2018run} 
  have been applied to models of active matter\cite{evans2018run,refractory},  predator-prey dynamics \cite{mercado2018lotka,toledo2019predator}, population dynamics \cite{ZRPSS,ZRPResetting},
 and stochastic processes \cite{lapeyre2019stochastic,gupta2018stochastic,basu2019long,basu2019symmetric} (see \cite{topical} for a  review, and references therein for more applications). For experimental realisations, see \cite{experimentalResetting}.\\

 It is natural to ask how the one-dimensional  picture of optimal resetting is deformed by the presence of a weak disorder in the environment. 
 In the presence of random forces, a diffusive random walk corresponds to the transverse coordinate 
 of the end-point of a directed polymer, described by the random-force model (or Larkin model).
 In this model of a directed polymer, random forces are transverse. Their amplitudes 
 depend only on the longitudinal coordinate (this coordinate is the directing coordinate and can be thought as time: the system is a directed polymer in 1+1 dimension). Moreover these amplitudes are independent and identically distributed.
 This model emerged as a model for pinning in superconductivity \cite{LarkinOriginal,LarkinReview,LeDoussalReview}, and its free-energy has been studied 
 for fixed and free boundary conditions in \cite{gorokhov1999exact,DotsenkoLarkin}.
 The linear structure of the contribution of the disorder to the 
 energy of the polymer makes the model the most elementary modification of the 
 diffusion by disorder. The model enjoys exact-solvability properties because it gives rise to Gaussian path integrals (for more general developments on path integrals for systems under resetting, see \cite{resettingPathIntegrals}).\\

 In Section 2, we will review the Larkin model in the absence of resetting, following the derivation of \cite{DotsenkoLarkin} (the relevant Gaussian integrals
 are exposed in Appendix A). We will work out the probability 
 density of the transverse position of the end-point of the polymer.  In Section 3 we will
 express the mean first-passage time at a fixed absorbing target, following the general arguments of \cite{rednerguide}, which are valid thanks to the 
 independence of the configurations of random forces in distinct intervals between resetting times. The optimal resetting rate 
 is defined as the rate that minimises this mean  first-passage time (which is known exactly in the absence of disorder). The  susceptibility of the optimal resetting rate 
  to disorder will be worked out as the quotient of two derivatives of the mean time. These two derivatives are obtained in closed form in Appendix B.



\section{Model and quantities of interest}
\subsection{The random-force model}
 Let us consider a directed polymer in $1+1$ dimension. An elastic line grows randomly on a plane. 
 The plane is endowed with Cartesian coordinates, and we think of $x$, the directing coordinate, as time.
   The configuration of the polymer is therefore decribed by the transverse displacement field
\begin{equation}
 \phi:  x\in [0,L] \mapsto \phi( x ) \in \mathbf{R}.
\end{equation}
  Let us assume the polymer start as a point-like object at the origin:
\begin{equation}
 \phi( 0 ) = 0.
\end{equation}
 Consider a disordered  environment described by a random potential  depending on the coordinate $x$ and the transverse displacement field.
 Moreover, let us assume that this potential corresponds to a random force which depends only on the coordinate $x$ and not on the current value 
 of the transverse displacement. Let us  denote this random force by $f(x)$. The corresponding random potential can therefore be expressed as $f(x)\phi(x)$.
   Let us denote the elasticity constant of the polymer by $c$. The energy of a configuration of the polymer at a fixed value $L$ of the directing coordinate 
 is the sum of the elastic energy and the potential energy:
\begin{equation}
 H[\phi,L,f]:= \int_0^L dx\left[   \frac{c}{2}\left( \frac{d\phi}{dx}(x)\right)^2 +  f(x)\phi(x)  \right].
\end{equation}
 The distribution of the random force field $f$ is assumed to be a centered Gaussian, with variance denoted by $u$. Denoting averages over disorder by 
 bars, we obtain
\begin{equation}\label{fLaw}
  \overline{f(x)} = 0,\;\;\;\;\;\overline{f(x)f(x')} = u\delta( x-x').
\end{equation}
 This choice of disordered potential is referred to as the  random-force model (or  Larkin model \cite{LarkinOriginal,LarkinReview,LeDoussalReview}). Setting 
 the parameter $u$ to zero we recover the free case, in which the end-point of the polymer is a diffusive Brownian random walker (the diffusion 
 constant will be expressed in terms of the temperature and elasticity constant in Eq. (\ref{D})).\\

\subsection{Partition function of a sector with fixed ends, for a given realisation of the random forces}

 Consider a given realisation of the force field $f$. 
 For a value $L$ of the directing coordinate, consider the sector consisting of all the transverse displacement fields such that the 
 end-point of the polymer is at a fixed value $y$ (in this sector $\phi(L)=y$). If thermal agitation is described by the parameter $\beta$, 
  this sector corresponds to the following Boltzmann weight: 
 \begin{equation}
Z[L,y;f] := \int_{\phi(0)=0}^{\phi(L)=y} [D\phi(x)]\exp\left(-\beta     H[\phi,L, f ] \right).
\end{equation}
  This functional integral is Gaussian and can therefore be explicitly evaluated. Let us follow the derivation 
 of \cite{DotsenkoLarkin} and impose the boundary conditions by changing function from the displacement field $\phi$ to the 
 field $\varphi$ defined by the linear shift:
\begin{equation}
 \phi(x) = \frac{x}{L}y + \varphi(x),\;\;\;\;\;{\mathrm{with}}\;\;\;\;\;\varphi(0) = \varphi(L) = 0.
\end{equation}

 \begin{equation}
\begin{split}
Z[L,y;f] &= \exp\left(-\beta c\frac{y^2}{L} -\beta  f(x) \int_0^L dx \frac{x}{L}y  \right) \\
 & \times\int_{\varphi(0)=0}^{\varphi(L)=0} 
[D\varphi(x)]\exp\left( -\beta  \left[\int_0^L dx\left(   \frac{c}{2} (\varphi'(x))^2  +  cy \varphi'(x) +  f(x)  \frac{x}{L}y  + f(x) \varphi(x) \right)    \right] \right).
\end{split}
\end{equation}
 Let us denote by $\varphi_q$ the configuration of the field $\varphi$ that makes the energy stationary, and by $\delta \varphi$ the fluctuations
 around this value:
\begin{equation}
 \varphi(x) =  \varphi_q(x) + \delta \varphi(x),
\end{equation}
 where $\varphi_q$ satisfies the equation
 \begin{equation}\label{stationary}
  c \varphi_q''(x) = f( x ),\;\;\;\;\;\;x\in [0,L].
\end{equation}
The fluctuations decouple in the following sense:
 \begin{equation}
\begin{split}
Z[L,y;f] =& \exp\left(-\beta c\frac{y^2}{2L}  -\beta y\int_0^L dx x f(x) \right)   \exp\left ( -\beta H[\varphi_q, f \varphi_q] \right)\\
 & \times\int_{\delta\varphi(0)=0}^{\delta\varphi(L)=0} [D \delta \varphi(x)]
\exp\left( -\beta  \int_0^L dx  \frac{c}{2} (\delta \varphi)'(x)^2  \right)\\
  &\times\exp\left(+\frac{y}{L}  \int_0^L dx( \varphi'_q(x) + (\delta \varphi)'(x) ) + \int_0^L dx ( c \varphi'_q(x)   (\delta \varphi)'(x) + f(x) \delta\varphi(x ) )  \right).
\end{split}
\end{equation}
 In the argument of the last exponential factor in the above equation, the first integral equals zero because of the boundary conditions,
 and the second integral equals zero because $\varphi_q$ statisfies the stationarity condition of Eq. (\ref{stationary}).
 Let us extend the transverse displacement $\varphi_q$ and force field $f$ to the interval $[-L,L]$, as odd functions. An odd function $g$ defined on $[-L,L]$ can be expanded as a Fourier series:
\begin{equation}\label{kmDef}
g_m:= \int_{-L}^{L} dx g(x) \sin( k_mx),\;\;\;\;{\mathrm{where}}\;\;\;\;k_m = \frac{m\pi}{L},\;\;\;\;m\in \mathrm{N},
\end{equation}
\begin{equation}
g( x ) = \frac{1}{L} \sum_{m=1}^\infty g_m \sin( k_m x ),\;\;\;\;\;\;x\in [-L,L].
\end{equation}
 The integrals needed in the expression of the energy are expressed as
\begin{equation} 
  \int_0^L  x f(x) dx = \frac{1}{L} \int_0^L x\sum_{m\geq 1} f_m \sin( k_m x) = \frac{1}{L}\sum_{m\geq 1} f_m \int_0^L x \sin( k_m x ) dx = - \frac{1}{L}\sum_{m\geq 1} (-1)^m \frac{f_m}{k_m}.
\end{equation}
\begin{equation} 
\begin{split}
H[\varphi_q, f \varphi_q] &= \frac{c}{2}\int_{-L}^L  dx \left[ (\varphi_q'(x))^2 - \varphi_q''(x)\varphi(x)\right]\\
                    & = -\frac{c}{4}\int_{-L}^L  dx (\varphi_q'(x))^2 \\
   &= -\frac{c}{4} \sum_{m\geq 1}\frac{k_m^2 \varphi_{qm}^2}{L^2} \int_{-L}^L  \cos^2(k_m x) dx\\
  & = -\frac{1}{4cL} \sum_{m\geq 1}\frac{f_m^2}{k_m^2}.
\end{split}
\end{equation}
where we used integration by parts and $-c k_m^2(\varphi_q)_m = f_m$, which is the $m$-th Fourier coefficient of  Eq. (\ref{stationary}).
 The partition fuction of the sector with transverse displacement $y$ at time $L$ therefore reads for a fixed realisation 
 of the random force field:
 \begin{equation}\label{partitionRes}
\begin{split}
Z[L,y;f] &= Z_0\exp\left(-\beta c\frac{y^2}{2L}  +\beta y  \frac{1}{L}\sum_{m\geq 1} (-1)^m \frac{f_m}{k_m}  + 
 \frac{\beta}{4cL} \sum_{m\geq 1}\frac{f_m^2}{k_m^2}\right),\\
 {\mathrm{with}}\;\;\; Z_0&:=  \int_{\delta\varphi(0)=0}^{\delta\varphi(L)=0} [D \delta \varphi(x)]
\exp\left( -\beta  \int_0^L dx  \frac{c}{2} (\delta \varphi)'(x)^2  \right).
\end{split}
\end{equation}
 The partition function of fluctuations with both ends fixed, denoted by $Z_0$, does not depend on the position $y$ of the end-point of the polymer.  In \cite{DotsenkoLarkin}, this result was used to study the free energy of the random-force model with fixed boundary condition. 
  We are going to use it to study the random transverve position of the end-point of the polymer instead.

\subsection{Probability law of the transverse position of the end-point}
  From Eq. (\ref{partitionRes}), we can derive the probability law of the random transverse transverse position $y$  of the end-point of the polymer.
  The normalisation factor we need  is the integral of the  expression  obtained in Eq. (\ref{partitionRes}) w.r.t. the coordinate $y$. This integral is Gaussian:  
\begin{equation}
\int_{-\infty}^{+\infty}Z[L,y;f] dy = Z_0 \sqrt{\frac{2\pi L}{\beta c}}
\exp\left( +\frac{L}{2\beta c}
\frac{\beta^2}{L^2}\left( \sum_{m=1}^\infty \frac{(-1)^m f_m}{k_m} \right)^2 \right)  \exp\left( \sum_{m=1}^\infty\frac{\beta f_m^2}{4cLk_m^2}\right).
\end{equation}  
 
 Let us denote by $P(y,t|y_0,t_0,[f])$ the probability density of the transverse  position of the end-point at coordinate $y$ at time $t$, conditional on the coordinate $y_0$ at time $t_0<t$, and on a particular realisation of forces $f$ (as the values of  the force $f$ at different times are independent, conditioning on the realisation $f$ is
 equivalent to conditioning on its restriction to the interval $[t_0,t]$).\\
  For a fixed realisation of forces, these notations yield
\begin{equation}\label{GaussianDeformed}
\begin{split}
P(y,L|0,0,[f])& := \frac{Z[L,y;f]}{\int_{-\infty}^{+\infty}Z[L,y;f] dy}\\
 &=  \sqrt{\frac{\beta c}{2\pi L}}\exp\left(  -\frac{\beta  c y^2}{2L} \right)
 \exp\left( y\frac{\beta}{L}\sum_{m=1}^\infty \frac{(-1)^m f_m}{k_m} -\frac{\beta}{2Lc}\left(\sum_{m=1}^\infty \frac{(-1)^m f_m}{k_m} \right)^2 \right).
\end{split}
\end{equation}
  The above expression is Gaussian in the linear combination of Fourier modes of the force field $f$.\\ 
   
 In the special case where  forces are identically zero (or equivalently where the amplitude $u$ is zero in Eq. (\ref{fLaw}))
  the model describes the free evolution of a directed elastic line. The probability law of the transverse position of the end-point becomes that of the position of an ordinary diffusive random walker in one dimension. Let us 
 use the symbol $t$ for the current value of the directing coordinate, and write Eq. (\ref{GaussianDeformed}) in the case of zero dirsorder: 
 \begin{equation}\label{diffusive}
P(y,t|0,0, [f=0]) = \frac{1}{\sqrt{4\pi Dt }}\exp\left(  -\frac{y^2}{4Dt} \right),
\end{equation}
where we read off  the diffusion constant $D$ in terms of the temperature and elastic constant of the problem as 
\begin{equation}\label{D}
 D:= \frac{1}{2 \beta c}.
\end{equation} 
 Turning on a disorder of  small amplitude $u>0$ allows non-zero configurations of the forces and  modifies the probability law of the transverse coordinate $y$,
 for each realisation of the forces, according to Eq. (\ref{GaussianDeformed}). Averaging over the disorder by 
 taking into account the Gaussian distribution of forces (Eq. (\ref{fLaw}))  should induce one-parameter deformations of all the properties of the diffusive random walk, including the optimal resetting rate.\\


 Let us denote by $\pi_L$ the probability density of the Fourier components of the forces (at a fixed amplitude $u>0$ of the disorder).
 We are interested in the   average over disorder  of the probability density of the position of the end-point of the polymer (conditional on the position $y_0$ at time $t_0<t$), 
 expressed as the following  integral:
\begin{equation}
\overline{P}(y,t|y_0,t_0) := \left(\prod_{n=1}^\infty\int_{-\infty}^{+\infty} df_n \pi_L(f_n) \right) P(y,t|y_0,t,[f]).
 \end{equation}
 The density $\pi_L$ is Gaussian (it is induced by the Gaussian distribution of the forces, Eq. (\ref{fLaw}), see Eq. (\ref{componentDensity}) in Appendix A). The integral over each 
  of the Fourier components is therefore Gaussian. After a calculation (worked out in Appendix A) we obtain 
 the following probability density of the transverse position of the end-point of the directed polymer at time $t$:
\begin{equation}\label{propagDensity}
\overline{P}(y,t|0,0)  =   \frac{1}{\sqrt{4\pi Dt\left( 1 + \epsilon t^2 \right) }}\exp\left(  -\frac{y^2}{4Dt \left( 1 + \epsilon t^2 \right)} \right),
 \end{equation}
with corrections w.r.t. the free diffusive case encoded by the parameter
\begin{equation}\label{epsilon}
 \epsilon: = \frac{\beta}{3c}u.
\end{equation} 
 Moreover, starting the process at time $t_0 <t$ with the condition $\phi(t_0) = y_0$, we can repeat 
 the above derivation using the fact that the forces $f_{[t_0,t]}$ have the same distribution as $f_{[0,t-t_0]}$,
 and obtain 
 \begin{equation}
\overline{P}(y,t|y_0,t_0)  =  \overline{P}(y -y_0,t -t_0| 0,0).  
 \end{equation}
 For our purposes, this result concludes the review of the Larkin model in the absence of resetting.

\section{The random-force model under resetting}

\subsection{First-passage time at a fixed target, averaged over disorder}

 Let us subject the directed polymer to stochatic resetting to a point-like configuration:
  at random times distributed according to a Poisson process with rate $r$, the previously-grown section of polymer is cut, and the position of the random walker (i.e. the transverse displacement of the polymer) 
 is reset to $0$. Morover, let us put  an absorbing wall at a fixed value $Y$ of the transverse coordinate: when the transverse coordinate of the end-point of the 
 directed polymer reaches the value $Y$, the process stops.  Let us denote by $\langle \overline{T}(Y,r,u)\rangle$ the average over disorder of the first-passage time of the end-point of the polymer
  at the transverse position $Y$.  In the case where $u=0$, the problem 
 reduces to ordinary one-dimensional Brownian motion (with diffusion constant $D$ defined in Eq. (\ref{D})). The mean first-passage time $\langle \overline{T}(Y,r,u=0)\rangle$ has been calculated, and found to exhibit a unique minimum for value of the resetting rate \cite{evans2011diffusion,evans2011optimal}, denoted by $r^\ast$. This 
 optimal value of the resetting rate depends only on the position $Y$ of the absorbing target, and on the diffusion constant (the expression of $r^ast$ is  reviewed at the beginning of Appendix B).\\

 For a non-zero value $u$ of the amplitude of the disorder, the mean first-passage time can again be expressed in integral form, as  the arguments leading to it are quite 
general \cite{rednerguide}. Let us denote by $S_{r,u}(Y,t|[f])$ the survival probability of the process until time $t$, 
 for a fixed realisation of forces on the interval $[0,t]$ (and an initial tranverse position $0$ at time $0$):
\begin{equation}
 S_{r,u}(Y,t|[f]) = \mathrm{Prob}\left( y<Y, \forall t' \in [0,t]|[f_{[0,t]}]\right).
\end{equation}
 Its average over disorder is denoted by $\overline{S_{r,u}}(Y,t)$.\\

The mean first-passage time  at $Y$ (averaged over disorder) is  the integral of  time against the decreasing rate of the survival probability, which can 
 be expressed in terms of the Laplace transform of the survival probability $\overline{S_{r,u}}$ by integrating by parts:
\begin{equation}\label{meanTime}
\langle \overline{T}(Y,r,u) \rangle= \int_0^{\infty} dt \, t \left( -\frac{\partial \overline{S_{r,u}}}{\partial t}( Y,t)\right) = \int_0^\infty \overline{S_{r,u}}( Y,t)dt =  \widetilde{\overline{S_{r,u}}}( Y,0),
\end{equation}
 where we denoted with a tilde the Laplace transform in the time variable:
\begin{equation}
 \tilde{g}(s) := \int_0^\infty dt\, e^{-st} g(t).
\end{equation}

\subsection{Renewal equation}

 On the other hand, the Laplace transform of the survival probability under resetting $\overline{S_{r,u}}$
  can be expressed in terms 
 of the Laplace transform of the survival probability in the ordinary process (with zero resetting rate and 
 a disorder of amplitude $u$), taken at the resetting rate. Consider a fixed realisation of the 
 random forces on the interval $[0,t]$. 
 Conditioning on the last resetting event in the interval $[0,t]$ induces a renewal equation:
\begin{equation}   
S_{r,u}(Y,t|[f]) = e^{-rt} S_{0,u}(Y,t|[f]) + r \int_0^t d\tau  e^{-r\tau}  S_{r,u}(Y,t-\tau|[f_{[0,t-\tau[}]) S_{0,u}(Y,\tau|[f_{[t-\tau,t]}]),
\end{equation}
where the first term corresponds to no resetting in the interval $[0,t]$ (hence the multiplicative factor $e^{-rt}$) and the second term to at least one resetting event 
 in this interval, the last of which occurs at $t-\tau$, for some $\tau$ in $[0,t]$.  Each factor in the integrand of the second term 
 is conditioned on the forces, but in the first factor the condition depends only on the forces on the interval $[0,t-\tau[$, and in the second factor the condition depends only on the forces on the interval $[t-\tau,t]$.\\

 To take the average of this renewal equation over disorder,
 let us use the independence of the two random configurations of forces $f_{[0,t-\tau[}$ and  $f_{[t-\tau,t]}$, 
 to write the average of the integrand as a product of averages. Moreover, the restriction $f_{[t-\tau,t]}$ has the same distribution  as $f_{[0,\tau]}$, hence
\begin{equation}   
\begin{split}
\overline{ S_{r,u}}(Y,t) &= e^{-rt} \overline{S_{0,u}}(Y,t) + r \int_0^t d\tau  e^{-r\tau} \overline{ S_{0,u}(Y,t-\tau) S_{0,u}(Y,\tau)}\\
 & = e^{-rt} \overline{S_{0,u}}(Y,t) + r \int_0^t d\tau  e^{-r\tau} \overline{ S_{r,u}}(Y,t-\tau)  \overline{S_{0,u}}(Y,\tau).
\end{split}
\end{equation}
 The Laplace transform w.r.t. $t$ yields
\begin{equation} 
\widetilde{\overline{ S_{r,u}}}(Y,s) = \widetilde{\overline{ S_{0,u}}}(Y,r+s) + r   \widetilde{\overline{ S_{0,u}}}(Y,r+s) \widetilde{\overline{ S_{r,u}}}(Y,s),
\end{equation}
 hence the averages over disorder of the survival probabilities satisfy the equation
 \begin{equation}\label{renewalLaplace}
\widetilde{\overline{ S_{r,u}}}(Y,s)
 = \frac{ \widetilde{\overline{ S_{0,u}}}(Y,r+s)}{ 1 - r   \widetilde{\overline{ S_{0,u}}}(Y,r+s)},
\end{equation}
 which reduces when $u=0$ to the known equation satisfied by the survival probability without disorder. To express the time 
 $\langle \overline{T}(Y,r,u) \rangle$, we see
 from Eqs  \ref{meanTime} and  \ref{renewalLaplace}  that it is enough  to evaluate the Laplace transform  of the survival probability $\overline{ S_{0,u}}$ in the process without resetting.\\

For the directed polymer in a fixed realisation of the forces, and with no resetting, let us denote 
 by $\phi_0(Y,t|f)$ the probability density of reaching the transverse displacement $Y$ for the first time at $t$ (after having started at transverse position $0$ at time $0$).
 This quantity describes the leaking of survival probability through the absorbing target, hence
\begin{equation}
 \phi_0( Y,t|f)= -\frac{\partial}{\partial t} S_{0,u}(Y,t|f).
\end{equation}
  Upon average over disorder and  Laplace transform, this yields the quantity we need in Eq. (\ref{renewalLaplace}):
\begin{equation}\label{StoPhi}
 \widetilde{\overline{\phi_0}}( Y,s) = 1 - s \widetilde{\overline{S_{0,u}}}(Y,s).
\end{equation}

 On the other hand, the density $\phi_0$ is related to the probability density of the position of the end-point of the polymer   
 by conditioning on the time $T$ at which the end-point   reaches the absorbing wall at  $Y$ 
 for the first time. Indeed we can write
\begin{equation}\label{aboveEq}
P(Y,t|0,0,[f])= \int_0^t dT \phi_0(Y,T| [f]) P( Y, t | Y, T, [f]). 
\end{equation}
 where the second factor in the integrand describes the return of the end-point to the transverse position $Y$ at $t$.
 The two factors in the integrand depend on the realisation of forces $f$ through their restrictions to the 
 intervals $[0,T[$ and $[T, t]$ respectively. As the forces at distinct times are independent and identically distributed, the average over disorder of 
 Eq. (\ref{aboveEq}) reads
\begin{equation}\label{convolo}
\begin{split}
\overline{P}(Y,t|0,0) &= \int_0^t dT \overline{\phi_0}(Y,T) \overline{P}( Y, t | Y, T )\\
 &= \int_0^t dT \overline{\phi_0}(Y,T) \overline{P}(Y, t- T | Y,0 )\\
 &= \int_0^t dT \overline{\phi_0}(Y,T) \overline{P}(0, t- T | 0,0 ),
\end{split}
\end{equation}
 where in the last step we used the fact the probability of return to the initial position in a fixed time
 is independent of the value of this initial position.
 Let us make the dependence on the amplitude of the disorder explicit by introducing the following 
  notation for the relevant Laplace transforms:
 \begin{equation}
 \mathcal{L}(y,r,u):= \int_0^\infty dt\, e^{-rt }\overline{P}(y, t | 0,0 ).\\
 \end{equation}

 The  Laplace transform  of Eq. (\ref{convolo}) w.r.t. the variable $t$ (taken at the value $r$) reads
 \begin{equation}\label{P0ToPhi}
   \mathcal{L}(Y,r,u)= \widetilde{ \overline{\phi_0}}(Y,r)  \mathcal{L}(0,r,u).
\end{equation}
 Combining Eqs (\ref{meanTime}, \ref{renewalLaplace}, \ref{StoPhi}, \ref{P0ToPhi}) yields
 \begin{equation}\label{Tused}
\langle \overline{T}(Y,r,u) \rangle= \frac{1}{r}\left(  1- \widetilde{ \overline{\phi_0}}(Y,r) \right)\frac{1}{\widetilde{ \overline{\phi_0}}(Y,r)} = 
\frac{1}{r}\left( \frac{\mathcal{L}(0,r,u)}{\mathcal{L}(Y,r,u)} -1 \right),
\end{equation}
 which is formally identical to the expression of the mean time to absorption in the absence of disorder
 in terms of the Laplace transform of the propagator of the process without resetting.\\

\subsection{Susceptibility to disorder  of the optimal resetting rate}
%
%
 
We would like to calculate the response rate of the optimal resetting rate to a disorder of infinitesimal amplitude.
  If we denote by $\rho(u)$ the optimal resetting rate for a small value of $u$ (at fixed values of the diffusion constant $D$, and fixed  position $Y$ of the target), so that  $\rho(0) = r^\ast$, and the 
 the desired susceptibility  is
\begin{equation}
 \frac{\delta r^\ast}{ \delta u} = \rho'( 0).
\end{equation}
 
 The optimality condition defining $\rho(u)$ takes the form
\begin{equation}\label{optimCond}
\frac{\partial}{\partial r} \langle \overline{T} (Y,r = \rho(u), u ) \rangle = 0.
\end{equation}

  
 The derivative $\rho'(0)$ can be calculated from the first-order term in the expansion of this optimality condition
 in powers of the amplitude of the disorder:
\begin{equation}
\frac{\partial}{\partial u}\left(  \frac{\partial}{\partial r} \langle \overline{T}(Y, r=\rho(u), u) \rangle \right)|_{u = 0} = 0.
\end{equation}
The susceptibility of the optimal rate to the disorder is therefore expressed as
 \begin{equation}\label{instructed}
\frac{ \delta r^\ast}{\delta u} = -   \left(  \frac{\partial^2 }{\partial r^2 } \langle T (Y, r^\ast, 0)\rangle \right)^{-1} \frac{\partial^2}{\partial r \partial u } \langle T (Y, r^\ast, 0)\rangle.
\end{equation}
 Thanks to the expression of the mean first-passage time in Eq. (\ref{Tused}), 
    the two second derivatives we are instructed to compute can be evaluated from a second-order Taylor expansion 
 of the Laplace transform of the  probability density of the end-point of the polymer 
 without resetting (Eq. (\ref{propagDensity})), around the values $r=r^\ast$ and $u = 0$. By dominated convergence,
  the terms we need can be obtained by expanding the integrands in the Laplace transforms 
 in powers of $r-r^\ast$ (up to order two) and $u$ (up to order one).
The derivation can be found in Appendix B. Thanks to the optimality condition satisfied by 
 $r^\ast$ in the absence of disorder, the susceptibility can be expressed in closed form in terms of 
  values of modified Bessel functions at the point $\gamma^\ast$ (Eq. (\ref{transc})). Numerically we find
\begin{equation}\label{deltarastdeltau}
\frac{ \delta r^\ast}{\delta u} 
\simeq   2.6394 \frac{\beta}{c r^\ast} \simeq  0.5197 \beta^2 Y^2 =:\zeta  \beta^2 Y^2.
\end{equation}
 The product $\beta^2 Y^2$  is the only monomial in the dimensionful parameters $\beta,c,Y$ with the correct unit, 
  because $u$ has dimension of the square of of a force divided by time.  
  The numerical factor is expressed in terms of the constant $\gammast$ only (see Eq. (\ref{coeffNum})). \\

\subsection{Discussion}

 The optimal resetting rate is therefore increased by  a small amplitude of disorder, and the effect grows quadratically with 
  the position of the target. Intuitively, this comes from the fact that long distances are probed at long times by the directed polymer, and the form of the 
   propagator in Ep. (\ref{propagDensity}) deviates more and more from the Gaussian form at long times. Increasing the disorder increases the variance of the position of the 
    end-point of the polymer at fixed time. A fixed resetting rate is therefore expected to allow for longer excursions in the wrong direction when the disorder is turned on. Increasing the   resetting rate from the value $r^\ast$ is therefore intuitively beneficial when optimising the mean first-passage time,  as it cuts off these excursions.   However,  the validity of the linear approximation for a fixed value of the disorder is compromised when the position of the target becomes too large.\\

 The above calculation  can be used for linear approximations  at finite values of the disorder  
  only if the value of the disorder is small enough for the value of the predicted first passage to be small enough 
    to be consistent  with the Gaussian approximation. As the susceptibility we calculated grows proportionally the square of the 
    position $Y$ of the target, the consistency of the approach  depends on the position of the 
     target (and on the temperature).\\

     The first-passage time $\langle\overline{T}(Y, r^\ast,0 )\rangle$, which is the order-zero approximation to the first passage time in the Larkin model, 
       should be in the validity zone of the diffusive approximation to the propagator displayed in Eq. (\ref{propagDensity}):\\
       \begin{equation}\label{bound1}
       \epsilon \ll \frac{1}{\langle\overline{T}(Y, r^\ast,0 )\rangle^2}.
       \end{equation}
       Using Eqs (\ref{exprTbar},\ref{transc})  the optimal first passage time reads as follows in terms of the position of the target:
         \begin{equation}
        \langle\overline{T}(Y, r^\ast,0 )\rangle = \frac{Y^2}{\gammast(2-\gammast) D}.
       \end{equation}
       The relevant order of magnitude for the disorder parameter is therefore proportional to the inverse  fourth power of  the position of the target:
       \begin{equation}
       \epsilon \ll  (\gammast(2-\gammast))^2\frac{D^2}{Y^4} \simeq   0.4194 \frac{D^2}{Y^4}.
       \end{equation}
       This bound does not depend on the numerical result of Eq. (\ref{deltarastdeltau}). To take the value of the susceptibility into account, we must
       impose the continuity condition
 \begin{equation}
  \frac{ \delta r^\ast}{\delta u} \delta u \ll r^\ast.
 \end{equation}
 Using the computed value of $\zeta$ together with the expression of the parameter $u$ is terms of $\epsilon$, as well as the expression of the diffusion constant (Eqs (\ref{deltarastdeltau},\ref{D},\ref{epsilon})), we obtain
  \begin{equation}
 \frac{ 3}{2D}\zeta  Y^2  \epsilon \ll   \frac{D(\gamma^\ast)^2}{Y^2},
 \end{equation}
  \begin{equation}
   \epsilon  \ll   \frac{2(\gamma^\ast)^2}{3\zeta}\frac{D^2}{Y^4} \simeq 3.2577 \frac{D^2}{Y^4}.
 \end{equation}
  This regularity condition is therefore automatically satisfied if $\epsilon$ satisfies the consistency condition  of Eq. (\ref{bound1}), based on the
   propagator. This condition therefore yields a condition on the position of the target at fixed disorder:
    \begin{equation}
    Y  \ll    (0.4194)^{1/4} \frac{\sqrt{D}}{\epsilon^{1/4}} = 0.8047  \frac{\sqrt{D}}{\epsilon^{1/4}}. 
    \end{equation}
   This bound can also be interpreted as  a lower bound on the temperature (an upper bound on $\beta$) for  fixed values of the position of the target and of the amplitude of the disorder.
      The higher the temperature is (at fixed elasticity constant and amplitude  of the disorder), the more important the effect of diffusion is, which is in favour
       of the perturbative approximation around the diffusive propagator.\\


    The approximation we made  is valid at short times, which as we have just seen induces upper  bounds  on the position 
      of the target, but it is bound to fail at large times, when the disorder becomes  dominant in the expression of the propagator $\overline{P}$ 
       displayed in Eq. (\ref{propagDensity}).
       An extreme case of inapplicability of our result to a first-order approximation of the optimal resetting rate is a case where the point 
       $(Y, \langle \overline{T}(Y,r^\ast,0)\rangle)$ is in the region  of space-time corresponding to the non-equilibrium 
        steady state of a model whose propagator is given by the large-time (or large-$\epsilon$) limit  of the propagator $\overline{P}$. Let us 
        follow the large-deviation reasoning of \cite{majumdar2015dynamical} to map this region (see also Sections 2.5 and 6 of \cite{topical} for a review). 
         Due to the structure of renewal equations,  
         the relaxation to a non-equilibrium steady state (at fixed resetting rate $r$) of this large-$\epsilon$ limit is governed by the following 
         integral against the exponential density, in which we changed the integration variable from $\tau$ in $[0,\tau]$ to $w=\tau/t$:
         \begin{equation}\label{saddle}
          \int_0^1 \frac{dw}{w^{3/2}}\exp\left(  -rt w - \frac{Y^2}{4D\epsilon t^3 w^3} \right) =:  \int_0^1 \frac{dw}{w^{3/2}}\exp\left( -t \Phi\left(w, \frac{Y}{t^2}\right) \right).
         \end{equation}
 We made the large-deviation function $\Phi$ appear, together with the natural scaling of the position w.r.t. time ($\xi := Y /t^2$):
 \begin{equation}
  \Phi\left( w,\xi\right) = rw + \frac{\xi^2}{4D\epsilon w^2}.
 \end{equation}
  The point of space-time  of coordinates $(Y,t)$  corresponds to a non-equilibrium steady state if the saddle point $w_\ast$ of the integral in Eq. (\ref{saddle})
   is in the open interval $]0,1[$ (otherwise the integral is dominated by the integrand at the final value and the regime is transient). The optimal 
    value of the integration variable satisfies 
    \begin{equation}
     0 = \frac{\partial \Phi}{\partial w}\left( w_\ast,\xi\right)  = r - \frac{3\xi^2}{4D\epsilon w_\ast^4}.
    \end{equation}
    For our local estimate of the optimal resetting rate to be relevant, we want to avoid the steady-state region in space-time, and therefore impose
    \begin{equation}
     w_\ast = \left( \frac{3\xi^2}{4D\epsilon r^{\ast}} \right)^{\frac{1}{4}}>1,\;\;\;\;\;{\mathrm{with}}\;\;\;\;\;\xi = \frac{Y}{\langle \overline{T}(Y,r^\ast,0)\rangle^2}.
    \end{equation}
     Using again the definition of the optimal resetting rate in terms of the position $Y$ and diffusion constant, we express this bound as
  an upper bound on the disorder:
  \begin{equation}
   \epsilon \ll  \frac{3}{4} \gammast^2 (2-\gammast)^4 \frac{D^2}{Y^4}\simeq 0.0520  \frac{D^2}{Y^4}.
   \end{equation}    
 This bound is stronger than the ones we obtained above. 
      Demanding that the point of space-time with coordinates $Y$ and the mean first-passage time be far from the steady-state region
        of the Larkin model  is therefore enough to imply that the amplitude of the disorder is small compared to the inverse square of the  unperturbed   optimal mean first passage  time.

\section{Conclusion}

 In this paper we have expressed the expectation value of the mean first-passage time to a fixed absorbing target (averaged over disorder)
 in the Larkin model of a directed polymer in random forces, subjected to stochastic resetting to a point-like configuration.
 The end-point of the polymer becomes an ordinary Brownian random walker when the 
  amplitude of the disorder goes to zero.
 The probability density of the transverse position of the end-point of the polymer can be averaged over disorder thanks to the Gaussian nature 
 of the involved integrals. Thanks to the independence properties of the random forces, the average over disorder of the mean first-passage time 
  to an absorbing target  has been expressed in closed form, in terms of the Laplace transform of the propagator of the process without resetting.
 To minimise this mean first-passage time,  we found it  beneficial on average to increase the resetting rate in presence of disorder  with  a small amplitude.\\

 This paper focused on the finite-time behaviour of the model. At large times the propagator deviates 
   from the diffusive form:
 thanks to  the equivalence $\overline{P}(y,t)\sim \exp\left( -y^2/(4D\epsilon t^3) \right)$ at large $t$,
 the Larkin model is of the class addressed in \cite{majumdar2015dynamical}. In particular, it exhibits a non-equilibrium 
 steady state on (a space-dependent) large time scale. In this paper we have been concerned in the regime of time in which 
 the effect of disorder is still small, and the behaviour of the end-point of the polymer is still close to a Brownian motion. 
 Technically, the involved expansions in powers of the amplitude of the disorder 
 are valid by dominated convergence, because they occur in the integrand of a Laplace transform.
  They yield the susceptibibility of the optimal resetting rate in one dimension to disorder. This susceptibility is proportional to the square 
 of the position of the target and to the inverse of the square of the temperature (for dimensional reasons), up to a positive coefficient expressed
 in closed form.\\

 Studying more complex disordered systems would typically involve the Laplace transform of the 
 probability distribution of the propagator of the system in the absence of resetting, taken at the resetting rate.
 In the case of the Matheron-de Marsily model 
 of a layered flow (with random velocities of the layers), special values of the Laplace transform of the propagator have been worked out 
 in the absence of resetting by path-integral methods \cite{LeDoussalMdm}. Extension of the calculation 
 of the Laplace transform to the resetting rate is usually a formidable task \cite{GeorgesReview}.
 However, in the case of the Matheron-de Marsily model, the transverse position 
 has been shown in \cite{majumdar2003persistence} to be a fractional Brownian motion, which thanks to the 
  large-deviation arguments of \cite{majumdar2015dynamical} induces the relaxation dynamics to 
 a non-equilibrium steady state under resetting.\\

%
%
%
%

 


\section*{Appendix A}

 From Eq. (\ref{fLaw}) we work out the mean and variance of the Fourier components of the 
  force field $f_{[-L,L]}$, the odd function obtained from $f_{0,L}$: 
\begin{equation}
 \overline{f_m} = 0,\;\;\;\; \overline{f_m^2} = \int_{-L}^L dx \int_{-L}^L dy \overline{f(x) f(y)} \sin(k_m x)\sin( k_m y) = 2uL.
\end{equation}
  Moreover, all the moments of higher order of  $f_m$ are zero, just as the moments of higher order of the forces.  Hence the probability density of $f_m$ (denoted by $\pi_L$, as it depends on the interval of time on which random forces 
 are studied) is the following centered Gaussian:
\begin{equation}\label{componentDensity}
\pi_L( f_m)  = \frac{1}{\sqrt{4\pi u L}} \exp\left(  -\frac{f_m^2}{4uL}  \right).
\end{equation}

 To average the probability density of the position of the end-point of the polymer, let us rewrite Eq. (\ref{GaussianDeformed}) as
 \begin{equation}
P(y,L|0,0,[f]) = \sqrt{\frac{\beta c}{2\pi L}}\exp\left(  -\frac{\beta  c y^2}{2L} \right)
 \exp\left(  \lambda y \vec{f}\cdot\vec{a}- \mu\left( \vec{f}\cdot\vec{a} \right)^2\right),
\end{equation}
 with the vector notations (using $k_m=m\pi/L$, Eq. (\ref{kmDef})):
 \begin{equation}
  \vec{f} := \sum_{n\geq 1} f_n\vec{e_n},\;\;\;\;\;\;
  \vec{a}= \frac{\sqrt{6}}{\pi}\sum_{n\geq 1} \frac{(-1)^n}{n}\vec{e_n},
 \end{equation}
 and the coefficients 
\begin{equation}
  \lambda :=  \frac{\beta}{\sqrt{6}},\;\;\;\;\mu := \frac{\beta L}{12 c}.
\end{equation}
The vector $\vec{a}$ is normalised.\\

The probability density of the position of the end-point of the polymer can be averaged over disorder by evaluating a Gaussian integral w.r.t. each of the Fourier components of the random force:
 \begin{equation}\label{GaussianProd}
 \begin{split}
\overline{P}(y,L|0,0) =& \left(\prod_{n=1}^\infty\int_{-\infty}^{+\infty} df_n \pi_L(f_n) \right) P(y,L|0,0,[f])  \\
 =& \sqrt{\frac{\beta c}{2\pi L}}\exp\left(  -\frac{\beta  c y^2}{2L} \right) 
 \left(\prod_{n=1}^\infty\int_{-\infty}^{+\infty} \frac{df_n}{\sqrt{4\pi u L}}  \right) \exp\left(  \lambda y \vec{f}\cdot\vec{a}- \mu\left( \vec{f}\cdot\vec{a} \right)^2 -     \frac{1}{4uL} \vec{f}\cdot\vec{f}\right).
 \end{split}
\end{equation}
 The quadratic terms in the Gaussian kernel are the sum of a multiple of the identity and 
 a multiple of the projector onto the normalised vector $\vec{a}$:
 \begin{equation}
 \mu\left( \vec{f}\cdot\vec{a} \right)^2 + \frac{1}{4uL} \vec{f}\cdot\vec{f}
  =: \vec{f}\cdot M \vec{f},
 \end{equation}
 where 
 \begin{equation}
 M_{ij} = \frac{1}{4uL} \delta_{ij} + \mu a_i a_j.
 \end{equation}
 We can invert $M$ as follows (where $I$ denotes the identity and $\pi_a$ denotes the projector onto the vector
 $\vec{a}$):
 \begin{equation}
 M^{-1}= 4uL\left( I + 4uL\mu \pi_a \right)^{-1}=
 4uL\left( I - \frac{4uL\mu}{1+4uL\mu} \pi_a \right).
 \end{equation}
 By Gram--Schmidt orthonormalisation, the determinant of $M$ reads
\begin{equation}
  \det M = (1+ 4uL \mu)\det\left( \frac{I}{4uL}\right) =  (1+ 4uL \mu) \prod_{n=1}^{\infty} \frac{1}{4uL}.
 \end{equation}
 The last factor in the above determinant  compensates the infinite product in the denominator in Eq. (\ref{GaussianProd}).
  The average over disorder therefore reads
\begin{equation}
\begin{split}
\overline{P}(y,L|0,0) & =\sqrt{\frac{\beta c}{2\pi L}}\exp\left(  -\frac{\beta  c y^2}{2L} \right) \frac{1}{\sqrt{1+ 4uL \mu}}\exp\left( \frac{\lambda^2y^2}{4} \vec{a}\cdot  M^{-1} \vec{a}\right)\\
&= \sqrt{\frac{\beta c}{2\pi L}}\exp\left(  -\frac{\beta  c y^2}{2L} \right) \frac{1}{\sqrt{1+ 4uL \mu}}\exp\left( \lambda^2y^2 uL \left(1- \frac{4uL\mu}{1+4uL\mu} \right)\right)\\
&= \sqrt{\frac{\beta c}{2\pi L}} \frac{1}{\sqrt{1+ \frac{\beta L^2 u}{3c} }}\exp\left( -\frac{\beta  c y^2}{2L}  +\frac{\beta^2}{6}y^2 uL  \frac{1}{1+\frac{\beta L^2 u}{3c}} \right)\\
&=  \sqrt{\frac{\beta c}{2\pi L\left(1+ \frac{\beta L^2 u}{3c} \right) }}
   \exp\left( -\frac{\beta  c y^2}{2L\left(1+ \frac{\beta L^2 u}{3c} \right)} \right).
\end{split}
 \end{equation}
 Using the expression of the diffusion coefficient introduced in Eq. (\ref{D}), we find
  the expression reported in Eq. (\ref{propagDensity}).

\section*{Appendix B}
\subsection*{Bessel-function identities}

We will repeatedly use the following identity (see Section 4.5 of \cite{handbook}):
\begin{equation}\label{Bessel}
\int_0^\infty dt \,t^{\nu - 1} \exp\left( -\frac{\alpha}{t} -\chi t \right) = 2 \left( \frac{\alpha}{\chi}\right)^{\nu/2} K_\nu( 2\sqrt{ \alpha\chi}),
\end{equation}
where $K_\nu$ denotes the modified Bessel function of the second kind of order $\nu$.
  The case $\nu = 1/2$  appears when calculating the Laplace
 transform of the probability density  of an ordinary random walk:
\begin{equation}\label{Kspec}
K_{1/2}( x ) = \sqrt{\frac{\pi}{2x}} e^{-x}.
\end{equation}
 This identity holds by continuity at $\alpha=0$ for $\nu >0$, because 
\begin{equation}\label{BesselEq}
  K_\nu(z) \underset{z\to 0}{\sim}  2^{\nu-1}\Gamma(\nu)z^{-\nu}.
 \end{equation}
  We will focus on the neighborhood of the optimal resetting rate in the absence of disorder. Hence all the values 
 of modified Bessel functions that we will need will be taken at the special point $\gamma^\ast$ 
  in terms of which the optimal resetting rate of the diffusive random walker can be expressed (see the next subsection for a review of the derivation, and Eq. (\ref{transc}) for the definition of $\gammast$). We will also use the special values $\Gamma(1/2) = \sqrt{\pi}$, $\Gamma( 3/2) = \sqrt{\pi}/2$, $\Gamma( 5/2) = 3\sqrt{\pi}/4$, $\Gamma( 7/2) = 15\sqrt{\pi}/8$.

\subsection*{Review of the optimal resetting rate in the absence of disorder}
At $u=0$ we are in the situation of the optimisation of resetting for diffusion in the absence of random forces. 
 The optimal rate   minimises:
\begin{equation}
\langle \overline{T}  (Y,r,u=0) \rangle= \frac{1}{r}\left( \frac{\mathcal{L}(0,r,u)}{\mathcal{L}(Y,r,u)} -1 \right).
\end{equation}
The quantities
 $\mathcal{L}(Y,r,0)$ and $\mathcal{L}(0,r,0)$  are just  Laplace transforms in time of the diffusive propagator of Eq. (\ref{diffusive}).
\begin{equation}\label{wo}
\begin{split}
\mathcal{L}(y,r,0) &= \int_0^\infty  P(y,t|0,0,[f=0])  e^{-rt} dt \\
&=  \int_0^\infty \frac{1}{\sqrt{4\pi Dt}}\exp\left( - \frac{y^2}{ 4Dt} -rt \right) dt\\
& = \frac{1}{\sqrt{4\pi D} }\times 2 \left( \frac{y^2}{4Dr}\right)^{1/4} K_{1/2}\left( r\sqrt{\frac{y}{D}}\right)\\
& = \frac{1}{ \sqrt{4Dr}} \exp\left(-y\sqrt{\frac{r}{D}} \right),
\end{split}
\end{equation}
 where we  used Eqs (\ref{Bessel},\ref{Kspec}).\\

  The optimal rate $r^\ast$  therefore minimises the quantity
\begin{equation}\label{exprTbar}
\langle \overline{T}  (Y,r,u=0) \rangle= \frac{1}{r}\left( \exp\left(\frac{Y}{\sqrt{D}}\sqrt{r}\right) - 1   \right)
\end{equation}
 in the variable $r$, for a constant position $Y$  of the absorbing target,
 which yields the condition
\begin{equation}
\frac{1}{2}\sqrt{\frac{r^\ast}{D}}Y = 1 - \exp\left( -\sqrt{\frac{r^\ast}{D}}Y \right).
\end{equation}
 The optimal rate is therefore  expressed in terms of the position of the absorbing target, the diffusion constant, and the solution $\gamma^\ast$ 
 to a transcendental equation:
\begin{equation}\label{transc}
 r^\ast = \frac{D(\gamma^\ast)^2}{Y^2},\;\;\;\;\;\;{\mathrm{with}}\;\;\;\;\;\frac{\gamma^\ast}{2} = 1 - \exp( -\gammast).
\end{equation}

Numerically $\gammast \simeq 1.5936$ (see Section 3 of \cite{topical} for an extensive review).
In the Taylor expansions around the optimal rate,  the  terms of order zero will involve the values of the Laplace transforms worked out in Eq. (\ref{wo})
 for $r=r^\ast$, which we can trade for polynomial expressions in the position of the absorbing target:
\begin{equation}\label{PYr}
 \mathcal{L}( Y,r^\ast, 0)= \frac{1}{2\sqrt{Dr^\ast}}\exp\left( -Y\sqrt{\frac{r^\ast}{D}}\right) = \frac{Y}{2 D\gammast}\exp( -\gammast),
\end{equation}

\begin{equation}\label{P0r}
\mathcal{L}( 0,r^\ast, 0)=  \frac{1}{2\sqrt{Dr^\ast}} = \frac{Y}{2 D\gammast}.
\end{equation}

 \subsection*{Taylor expansions}

 To evaluate the expressions in Eqs (\ref{Tused},\ref{instructed}),  we need a Taylor expansion of the Laplace transform
 $\mathcal{L}(y,r^\ast,u)$  around $(r=r^\ast, u= 0)$, at order two in the rate and order one 
  in the amplitude of the disorder. With the notations for the diffusion constant $D$ and noise parameter $\epsilon$ introduced in Eqs \ref{D} and \ref{epsilon}, we write
\begin{equation}
\mathcal{L}(y,r^\ast + h, u = 3c\beta^{-1}\epsilon) = \int_0^\infty dt \frac{1}{\sqrt{4\pi D t \left(  1 + \epsilon t^2  \right)}}
 \exp\left( -(r^\ast + h)t - \frac{y^2}{4Dt\left(  1 + \epsilon t^2  \right)} \right),
\end{equation}
  We will specialise the result to the values  $0$ and $Y$ of the transverse coordinate $y$.\\


The first-order terms in the noise parameter $\epsilon$ are extracted from
\begin{equation}
\frac{1}{\sqrt{\left(  1 + \epsilon t^2  \right)}} = 1 - \frac{t^2}{2}\epsilon + o(\epsilon),
\end{equation}
\begin{equation}
\begin{split}
  \exp\left( - \frac{y^2}{4Dt\left(  1 + \epsilon t^2  \right)} \right) &=  \exp\left( - \frac{y^2}{4Dt} \left(  1 - \epsilon t^2  + o(\epsilon)\right)\right)\\
&=\exp\left( - \frac{y^2}{4Dt} \right) \left( 1  +    \frac{y^2 t}{4D}\epsilon  + o(\epsilon )\right)\\
\end{split}
\end{equation}

The terms of order one and two in $h$ come only from the Taylor expansion of the exponential function in the factor $e^{-ht}$, present in both integrands:
\begin{equation}
 e^{-(r^\ast + h)t} = e^{-r^\ast t}\left( 1 - th + \frac{t^2}{2} h^2 + o(h^2)\right).
\end{equation}

\begin{equation}
\begin{split}
\mathcal{L}(y,r^\ast + h, u = 3c\beta^{-1}\epsilon) = & \int_0^\infty dt \frac{1}{\sqrt{4\pi D t}}\exp\left( - r^\ast t - \frac{y^2}{4Dt} \right)\left( 1 - th + \frac{t^2}{2} h^2 + o(h^2)\right) \\
 &\times \left( 1  +    \frac{y^2 t}{4D}\epsilon  + o(\epsilon )\right)\times \left( 1 - \frac{t^2}{2}\epsilon + o(\epsilon)\right)\\
= & \int_0^\infty dt \frac{1}{\sqrt{4\pi D t}}\exp\left( - r^\ast t - \frac{y^2}{4Dt} \right)\left( 1 - th + \frac{t^2}{2} h^2 + o(h^2)\right) \\
  &\times \left( 1  +    \frac{y^2 t}{4D}\epsilon   - \frac{t^2}{2}\epsilon + o(\epsilon) \right)\\
=: &  \mathcal{L}( y,r^\ast, 0) +  \kappa_{10}(y) h +  \kappa_{01}(y)\epsilon +   \kappa_{11}(y)h\epsilon + \kappa_{20} h^2 + \dots,
\end{split}
\end{equation}
where we can read off the expression of the coefficients in the Taylor expansions in integral form, and apply the identity of Eq. (\ref{Bessel}) to each term. 
\begin{equation}
\begin{split}
 \kappa_{10} (y) &= -\frac{1}{\sqrt{4\pi D}} \int_0^\infty dt  t^{1/2} \exp\left( -r^\ast t -\frac{y^2}{4Dt}\right)\\
 & =  -\frac{1}{\sqrt{4\pi D}} \times 2 \left(   \frac{y^2}{4Dr^\ast}\right)^{3/4} K_{3/2}\left( y \sqrt{\frac{r^\ast}{D}}\right).
\end{split}
\end{equation}
\begin{equation}
\begin{split}
 \kappa_{01}( y ) &= \frac{y^2}{4D\sqrt{4\pi D}} \int_0^\infty dt t^{1/2} \exp\left( -r^\ast t -\frac{y^2}{4Dt}\right)  -\frac{1}{2\sqrt{4\pi D}} \int_0^\infty dt t^{3/2} \exp\left( -r^\ast t -\frac{y^2}{4Dt}\right)\\
  & = \frac{y^2}{4D\sqrt{4\pi D}}  \times2 \left(  \frac{y^2}{4 Dr^\ast} \right)^{3/4} K_{3/2}\left( y\sqrt{\frac{r}{D}}\right)  
-\frac{1}{2\sqrt{4\pi D}}\times 2  \left(  \frac{y^2}{4Dr^\ast } \right)^{5/4}  K_{5/2}\left( y \sqrt{\frac{r^\ast}{D}}\right),
\end{split}
\end{equation}
\begin{equation}
\begin{split}
 \kappa_{11}( y )   &= - \frac{y^2}{4D\sqrt{4\pi D}} \int_0^\infty dt t^{3/2} \exp\left( -r^\ast t -\frac{y^2}{4Dt}\right) 
      +  \frac{1}{2\sqrt{4\pi D}} \int_0^\infty dt t^{5/2} \exp\left( -r^\ast t -\frac{y^2}{4Dt}\right)\\
 &=- \frac{y^2}{4D\sqrt{4\pi D}}\times 2 \left(   \frac{y^2}{4Dr^\ast}\right)^{5/4} K_{5/2}\left( y \sqrt{\frac{r^\ast}{D}}\right)
   +  \frac{1}{2\sqrt{4\pi D}} \times 2 \left(   \frac{y^2}{4Dr^\ast}\right)^{7/4} K_{7/2}\left( y \sqrt{\frac{r^\ast}{D}}\right).
\end{split}
\end{equation}
\begin{equation}
\begin{split}
 \kappa_{20}  (y)  &=  \frac{1}{2\sqrt{4\pi D}} \int_0^\infty dt t^{3/2} \exp\left( -r^\ast t -\frac{y^2}{4Dt}\right)\\
   &=  \frac{1}{2\sqrt{4\pi D}} \times 2 \left(   \frac{y^2}{4Dr^\ast}\right)^{5/4} K_{5/2}\left( y \sqrt{\frac{r^\ast}{D}}\right).
\end{split}
\end{equation}
 
The argument  of each of the modified Bessel functions is $y\sqrt{r^\ast/D}$ in  the above expressions, which for $y=Y$ reduces to 
 the constant $\gamma^\ast$, defined in Eq. (\ref{transc}). Moreover, this definition can be used to  eliminate the 
 symbol $r^\ast$ from the above expression if $y=Y$. This yields the following polynomial expressions in the position $Y$ of the absorbing target:
\begin{equation}
\frac{Y^2}{4D r^\ast} = \frac{Y^4}{4\gammast^2 D^2}.
 \end{equation}
 The values we will need are therefore obtained by substituting $Y$ and $0$ to the variable $y$.
\begin{equation}
 \kappa_{10} (Y)   =  -\frac{1}{\sqrt{4\pi D}} \times 2 \left(   \frac{Y^4}{4D^2 \gammast^2}\right)^{3/4} K_{3/2}\left( \gammast \right),
\end{equation}

\begin{equation}
 \kappa_{01}( Y ) = \frac{Y^2}{4D\sqrt{4\pi D}}  \times2 \left(    \frac{Y^4}{4D^2 \gammast^2}  \right)^{3/4} K_{3/2}\left( \gammast \right)  
-\frac{1}{\sqrt{4\pi D}}  \left(   \frac{Y^4}{4D^2 \gammast^2}    \right)^{5/4}  K_{5/2}\left( \gammast \right),
\end{equation}
\begin{equation}
 \kappa_{11}( Y)  =- \frac{Y^2}{4D\sqrt{4\pi D}}\times 2 \left(    \frac{Y^4}{4D^2 \gammast^2}    \right)^{5/4} K_{5/2}\left( \gammast \right)
   +  \frac{1}{2\sqrt{4\pi D}} \times 2 \left(  \frac{Y^4}{4D^2 \gammast^2}  \right)^{7/4} K_{7/2}\left( \gammast \right),
\end{equation}
\begin{equation}
 \kappa_{20}  (Y) =  \frac{1}{2\sqrt{4\pi D}} \times 2 \times    \left(    \frac{Y^4}{4D^2 \gammast^2}    \right)^{5/4}      K_{5/2}\left( \gammast \right).
\end{equation}
To obtain the needed values at $y=0$, let us  use the equivalents of modified Bessel functions  in Eq. (\ref{BesselEq}). Again let us use the definition of the 
 optimal resetting rate (Eq. (\ref{transc})) to make the dependence on $Y$ explicit.  We are going to encounter
 limits of the form  
\begin{equation}
 \lim_{y\to 0}\left(2  \left(\frac{y^2}{4Dr^\ast}\right)^{\nu/2} K_{\nu}\left( y \sqrt{\frac{r^\ast}{D}}\right) \right)= \frac{\Gamma(\nu/2)}{(r^\ast)^\nu} =
 \frac{\Gamma(\nu/2)}{D^\nu (\gammast)^{2\nu}}Y^{2\nu}.
\end{equation}
The needed coefficients at $y=0$  therefore read:
\begin{equation}
\begin{split}
 \kappa_{10} (0) &  =  \lim_{y\to 0}\left(     -\frac{1}{\sqrt{4\pi D}} \times 2 \left(   \frac{y^2}{4Dr^\ast}\right)^{3/4} K_{3/2}\left( y \sqrt{\frac{r^\ast}{D}}\right)      \right)\\
&=   -\frac{1}{\sqrt{4\pi D}}\frac{\Gamma(3/2)}{D^{3/2} (\gammast)^3}Y^3,
\end{split}
\end{equation}
\begin{equation}
\begin{split}
 \kappa_{01}( 0 ) &=  \lim_{y\to 0}\left(  \frac{y^2}{4D\sqrt{4\pi D}}  \times2 \left(  \frac{y^2}{4 Dr^\ast} \right)^{3/4} K_{3/2}\left( y\sqrt{\frac{r^\ast}{D}}\right)  
-\frac{1}{\sqrt{4\pi D}}  \left(  \frac{y^2}{4Dr^\ast } \right)^{5/4}  K_{5/2}\left( y \sqrt{\frac{r^\ast}{D}}\right) \right) \\
&=  -\frac{1}{2\sqrt{4\pi D}} \frac{\Gamma(5/2)}{D^{5/2} (\gammast)^5}Y^5,
\end{split}
\end{equation}
\begin{equation}
 \kappa_{11}( 0 )   = 
    \frac{1}{2\sqrt{4\pi D}}\frac{\Gamma(7/2)}{D^{7/2} (\gammast)^7}Y^7,
\end{equation}
\begin{equation}
 \kappa_{20}  (0)  
   =  \frac{1}{2\sqrt{4\pi D}}\frac{\Gamma(5/2)}{D^{5/2} (\gammast)^5}Y^5.
\end{equation}

We are therefore instructed to extract the relevant corrections from:
\begin{equation}\label{exprTuh}
 \langle \overline{T}( Y,r^\ast + h, \epsilon) \rangle =  \frac{1}{r^\ast + h}\left( \frac{   \mathcal{L}( 0,r^\ast, 0) +  \kappa_{10}(0) h +  \kappa_{01}(0)\epsilon +   \kappa_{11}(0)h\epsilon + \kappa_{20}(0) h^2 + \dots}{  \mathcal{L}( Y,r^\ast, 0)+  \kappa_{10}(Y) h +  \kappa_{01}(Y)\epsilon +   \kappa_{11}(Y)h\epsilon + \kappa_{20}(Y) h^2 + \dots }- 1\right).
\end{equation}
Factorising the dominant terms yields:
\begin{equation}\label{exprTuh}
\begin{split}
 \langle \overline{T}(Y, r^\ast + h, \epsilon) \rangle &=  \frac{1}{r^\ast}\left( 1 - \frac{1}{r^\ast} h+ \frac{1}{(r^\ast)^2} h^2+ o(h^2) \right)\\
  & \times \left( \frac{   \mathcal{L}( 0,r^\ast, 0)+  \kappa_{10}( 0 ) h +  \kappa_{01}(0)\epsilon +   \kappa_{11}(0)h\epsilon + \kappa_{20}(0) h^2 + \dots}{\mathcal{L}( Y,r^\ast, 0)  +  \kappa_{10}(Y) h +  \kappa_{01}(Y)\epsilon +   \kappa_{11}(Y)h\epsilon + \kappa_{20}(Y) h^2 + \dots }- 1\right)\\
& = \frac{1}{r^\ast}\left( 1 - \frac{1}{r^\ast} h+ \frac{1}{(r^\ast)^2} h^2+ o(h^2) \right)\\
& \times \left( \frac{   \mathcal{L}( 0,r^\ast, 0)}{ \mathcal{L}( Y,r^\ast, 0)  }   \times 
\frac{ 1 +  L_{10}( 0)  h +  L_{01}(0)\epsilon +   L_{11}(0)h\epsilon + L_{20}(0) h^2 + \dots}{  1 +  L_{10}(Y) h +  L_{01}(Y)\epsilon +   L_{11}(Y)h\epsilon 
 + L_{20}(Y) h^2 + \dots }- 1\right),\\
\end{split}
\end{equation}
with the notations
\begin{equation}
 L_{\delta}( y) = \frac{K_\delta(y) }{\mathcal{L}( y,r^\ast, 0)},
\end{equation}
for every pair of indices $\delta$.\\

The explicit values, using Eqs (\ref{PYr},\ref{P0r}) are as follows:
\begin{equation}
\begin{split}
 L_{10}( Y) &=   -\frac{1}{\sqrt{4\pi D}} \times 2 \left(   \frac{Y^4}{4D^2 \gammast^2}\right)^{3/4} K_{3/2}\left( \gammast \right) \frac{2D\gammast e^\gammast}{Y} = -\frac{ K_{3/2}\left( \gammast \right) e^\gammast}{\sqrt{2\pi\gammast}} \frac{Y^2}{D},\\
L_{01}(Y) &= \left( \frac{ K_{3/2}\left( \gammast \right) e^\gammast}{4\sqrt{2\pi\gammast}} 
-\frac{1}{4\sqrt{2\pi }}     \frac{K_{5/2}\left( \gammast \right) e^\gammast  }{(\gammast)^{3/2}}  
   \right)\frac{Y^4}{D^2},\\
 L_{11}( Y) &=  -\frac{Y^2}{2D}\frac{1}{4\sqrt{2\pi }} \frac{K_{5/2}(\gammast) e^\gammast  }{(\gammast)^{3/2}}\frac{Y^4}{D^2}
  +  \frac{1}{2\sqrt{4\pi D}} \times 2 \left(  \frac{Y^4}{4D^2 \gammast^2}  \right)^{7/4} K_{7/2}\left( \gammast \right) 
     \times\frac{2D\gammast e^\gammast}{Y} \\
 &= \left( - \frac{K_{5/2}(\gammast) e^\gammast  }{8\sqrt{2\pi }(\gammast)^{3/2}} +
\frac{ K_{7/2}(\gammast) e^\gammast}{8\sqrt{2\pi}(\gammast)^{5/2}} 
     \right)\frac{Y^6}{D^3},\\
 L_{20}(Y) &=  \frac{K_{5/2}(\gammast) e^\gammast }{4\sqrt{2\pi }(\gammast)^{3/2}} \frac{Y^4}{D^2}.
\end{split}
\end{equation}

\begin{equation}
 L_{10}( 0) = 
    -\frac{\Gamma(3/2)}{\sqrt{\pi} ( \gammast)^2 } \frac{Y^2}{D},\;\;\;\;\;\;\;
L_{01}(0) =  -L_{20}(0) =-\frac{\Gamma(5/2)}{2\sqrt{\pi} ( \gammast)^4 } \frac{Y^4}{D^2},\;\;\;\;\;\;\;
 L_{11}(0) =    \frac{\Gamma(7/2)}{2\sqrt{\pi} ( \gammast)^6} \frac{Y^6}{D^3},\;\;\;\;\;\;\;
\end{equation}

\begin{equation}
\begin{split}
\frac{ 1 +  L_{10}( 0)  h +  L_{01}(0)\epsilon +   L_{11}(0)h\epsilon + L_{20}(0) h^2 + \dots}{  1 +  L_{10}(Y) h +  L_{01}(Y)\epsilon +   L_{11}(Y)h\epsilon 
 + L_{20}(Y) h^2 + \dots }& = 1 +\\
 & ( L_{10}( 0) -  L_{10}(Y)   ) h \\
&+ (L_{01}(0) - L_{01}(Y)   ) \epsilon\\
 & + ( L_{11}(0) - L_{11}(Y)  - L_{01}(0)  L_{10}(Y)  - L_{10}( 0 ) L_{01}(Y) ) h\epsilon \\
 &+ (  L_{20}(0) - L_{20}(Y)  -  L_{10}( 0)  L_{10}(Y)  +  L_{10}(Y)^2 )h^2 + \dots\\
&=: 1 + M_{10} h + M_{01}\epsilon + M_{11}h\epsilon + M_{20} h^2 +\dots\\
\end{split}
\end{equation}

\begin{equation}\label{exprTuh}
\begin{split}
 \langle \overline{T}(Y, r^\ast + h, \epsilon ) \rangle 
& = \frac{1}{r^\ast}\frac{\mathcal{L}( 0,r^\ast, 0)}{\mathcal{L}( Y,r^\ast, 0)}\left( 1 - \frac{1}{r^\ast} h+ \frac{1}{(r^\ast)^2} h^2+ o(h^2) \right)\\\
& \times \left( 1- \frac{\mathcal{L}( Y,r^\ast, 0)}{\mathcal{L}( 0,r^\ast, 0)} + M_{10} h + M_{01}\epsilon + M_{11}h\epsilon + M_{20} h^2 +\dots\right),\\
& =: \frac{1}{r^\ast}\frac{\mathcal{L}( 0,r^\ast, 0)}{\mathcal{L}( Y,r^\ast, 0)}\left(  1- \frac{\mathcal{L}( Y,r^\ast, 0)}{\mathcal{L}( 0,r^\ast, 0)} +  \tau_{01} \epsilon + \tau_{10} h +  \tau_{11}h\epsilon + \tau_{20} h^2 +\dots\right)\\
 & =  \langle \overline{T}(Y, r^\ast, 0 ) \rangle +   \frac{1}{r^\ast}\frac{\mathcal{L}( 0,r^\ast, 0)}{\mathcal{L}( Y,r^\ast, 0)} \left(  \tau_{01} \epsilon + \tau_{10} h + \tau_{11} h\epsilon + \tau_{20} h^2 +\dots   \right),
\end{split}
\end{equation}
with 
\begin{equation}
 \tau_{01}= M_{01},
\end{equation}
\begin{equation}
  \tau_{10} = M_{10} - \frac{1}{r^\ast} \left(  1- \frac{\mathcal{L}( Y,r^\ast, 0)}{\mathcal{L}( 0,r^\ast, 0)}\right),
\end{equation}
\begin{equation}
 \tau_{11} = - \frac{1}{r^\ast}M_{01} + M_{11},
\end{equation}
\begin{equation}
  \tau_{20} = \frac{1}{(r^\ast)^2}\left(  1- \frac{\mathcal{L}( Y,r^\ast, 0)}{\mathcal{L}( 0,r^\ast, 0)}\right) + M_{20} - \frac{1}{r^\ast}   M_{10}.
\end{equation}

 The stationarity condition satisfied by the rate $r^\ast$ implies  $\tau_{10}=0$:
\begin{equation}
   M_{10} =  L_{10}( 0) -  L_{10}(Y)= \frac{1}{r^\ast} \left(  1- \frac{\mathcal{L}( Y,r^\ast, 0)}{\mathcal{L}( 0,r^\ast, 0)}\right).
\end{equation}
It gives rise to the following simplifications:
\begin{equation}
 \tau_{20}= M_{20}.
\end{equation}
\begin{equation}
\begin{split}
   M_{20} &= L_{20}(0) - L_{20}(Y) +  L_{10}(Y)(  L_{10}(Y)  -  L_{10}( 0) )\\
 & =  L_{20}(0) - L_{20}(Y) -  \frac{ L_{10}(Y)}{r^\ast} \left(  1- \frac{\mathcal{L}( Y,r^\ast, 0)}{\mathcal{L}( 0,r^\ast, 0)}\right)\\
 &=  L_{20}(0) - L_{20}(Y) -  \frac{ L_{10}(Y)}{r^\ast} \left(  1- e^{-\gammast}\right)\\
 & =  L_{20}(0) - L_{20}(Y) -  \frac{\gammast}{2}\frac{ L_{10}(Y)}{r^\ast}\\
& =  L_{20}(0) - L_{20}(Y) -   L_{10}(Y) \frac{Y^2}{2D\gammast}\\
& = \left( \frac{\Gamma(5/2)}{2\sqrt{\pi} ( \gammast)^4}   - \frac{K_{5/2}(\gammast) e^\gammast }{4\sqrt{2\pi }(\gammast)^{3/2}} +  \frac{K_{3/2}(\gammast) e^{\gammast}}{2\sqrt{2\pi}\gammast^{3/2}}\right)\frac{Y^4}{D^2}\\
& = \left( \frac{\Gamma(5/2)}{2 ( \gammast)^4}   - \frac{K_{5/2}(\gammast) e^\gammast }{4\sqrt{2 }(\gammast)^{3/2}} +  \frac{K_{3/2}(\gammast) e^{\gammast}}{2\sqrt{2}\gammast^{3/2}}\right)\frac{Y^4}{\sqrt{\pi}D^2}\\
\end{split}
\end{equation}

The desired susceptibility is therefore obtained as
\begin{equation}\label{needed}
\begin{split}
\frac{ \delta r^\ast}{\delta u} &= - \frac{\beta}{3c}  \frac{ \tau_{11} }{2 \tau_{20}} \\
&= +\frac{\beta}{6c r^\ast}\frac{M_{01} - r^\ast  M_{11}}{ M_{20}}. \\
\end{split}
\end{equation}

Moreover, the quantities  $M_{01}$ and $r^\ast M_{11}$  also carry a dimensionful factor of $Y^4/D^2$:
\begin{equation}
\begin{split}
   M_{01} &= L_{01}(0) - L_{01}(Y)\\
 & =  \left( -\frac{\Gamma(5/2)}{2\sqrt{\pi} ( \gammast)^4 } - \frac{ K_{3/2}\left( \gammast \right) e^\gammast}{4\sqrt{2\pi\gammast}} 
+\frac{1}{4\sqrt{2\pi }}     \frac{K_{5/2}( \gammast) e^\gammast  }{(\gammast)^{3/2}}  
   \right)\frac{Y^4}{D^2}\\
& =  \left( -\frac{3\sqrt{\pi}}{8( \gammast)^4 } - \frac{ K_{3/2}\left( \gammast \right) e^\gammast}{4\sqrt{2\gammast}} 
+   \frac{K_{5/2}( \gammast) e^\gammast  }{4\sqrt{2 }(\gammast)^{3/2}}  
   \right)\frac{Y^4}{\sqrt{\pi}D^2}\\
\end{split}
\end{equation}

\begin{equation}
\begin{split}
  r^\ast M_{11} =& r^\ast\left[   L_{11}(0) - L_{11}(Y)  - L_{01}(0)  L_{10}(Y)  - L_{10}( 0 ) L_{01}(Y)   \right]\\
  = &  \frac{\gammast^2 D}{Y^2}[ \frac{\Gamma(7/2)}{2\sqrt{\pi} ( \gammast)^6} \frac{Y^6}{D^3} 
 -  \left( - \frac{K_{5/2}(\gammast) e^\gammast  }{8\sqrt{2\pi }(\gammast)^{3/2}} +
\frac{ K_{7/2}(\gammast) e^\gammast}{8\sqrt{2\pi}(\gammast)^{5/2}} 
     \right)\frac{Y^6}{D^3}\\
&-   \frac{\Gamma(5/2)}{2\sqrt{\pi} ( \gammast)^4 } \frac{Y^4}{D^2} \times \frac{ K_{3/2}\left( \gammast \right) e^\gammast}{\sqrt{2\pi\gammast}} \frac{Y^2}{D}
+\frac{\Gamma(3/2)}{\sqrt{\pi} ( \gammast)^2 } \frac{Y^2}{D} \times   \left( \frac{ K_{3/2}\left( \gammast \right) e^\gammast}{4\sqrt{2\pi\gammast}} 
-\frac{1}{4\sqrt{2\pi }}     \frac{K_{5/2} e^\gammast  }{(\gammast)^{3/2}}  
   \right)\frac{Y^4}{D^2} ]\\
=&  \frac{\gammast^2 Y^4}{\sqrt{\pi}D^2 }[ \frac{\Gamma(7/2)}{ 2( \gammast)^6}
  + \frac{K_{5/2}(\gammast) e^\gammast  }{8\sqrt{2 }(\gammast)^{3/2}} -
\frac{ K_{7/2}(\gammast) e^\gammast}{8\sqrt{2}(\gammast)^{5/2}}\\
&-   \frac{\Gamma(5/2)}{ 2( \gammast)^4 } \times \frac{ K_{3/2}\left( \gammast \right) e^\gammast}{\sqrt{2\pi\gammast}}  
  +\frac{\Gamma(3/2)}{ ( \gammast)^2 }   \times   \left( \frac{ K_{3/2}\left( \gammast \right) e^\gammast}{4\sqrt{2\pi\gammast}} 
  -  \frac{K_{5/2}\left( \gammast \right) e^\gammast  }{4\sqrt{2\pi }(\gammast)^{3/2}}  
   \right) ]\\
=&  \frac{ Y^4}{\sqrt{\pi}D^2 }[ \frac{15\sqrt{\pi}}{ 16( \gammast)^4}
  + \frac{K_{5/2}(\gammast) e^\gammast  }{8\sqrt{2 }(\gammast)^{-1/2}} -
\frac{ K_{7/2}(\gammast) e^\gammast}{8\sqrt{2}(\gammast)^{1/2}}\\
&-   \frac{3\sqrt{\pi}}{8 ( \gammast)^2} \times \frac{ K_{3/2}\left( \gammast \right) e^\gammast}{\sqrt{2\pi\gammast}}  
  + \frac{\sqrt{\pi}}{2} \times   \left( \frac{ K_{3/2}\left( \gammast \right) e^\gammast}{4\sqrt{2\pi\gammast}} 
  -  \frac{K_{5/2}\left( \gammast \right) e^\gammast  }{4\sqrt{2\pi }(\gammast)^{3/2}}  
   \right) ]\\
=&  \frac{ Y^4}{\sqrt{\pi}D^2 }[ \frac{15\sqrt{\pi}}{ 16( \gammast)^4}
  + \frac{K_{5/2}(\gammast) e^\gammast  }{8\sqrt{2 }(\gammast)^{-1/2}} -
\frac{ K_{7/2}(\gammast) e^\gammast}{8\sqrt{2}(\gammast)^{1/2}}\\
&-    \frac{3 K_{3/2}\left( \gammast \right) e^\gammast}{8\sqrt{2} (\gammast)^{5/2}}  
  + \frac{1}{8\sqrt{2}}  \times   \left( \frac{ K_{3/2}\left( \gammast \right) e^\gammast}{\sqrt{\gammast}} 
  -  \frac{K_{5/2}\left( \gammast \right) e^\gammast  }{(\gammast)^{3/2}}  
   \right) ]
\end{split}
\end{equation}

 All terms in the numerator and denominator of the fraction needed in Eq. (\ref{needed}) contain factor of $Y^4/(\sqrt{\pi}D^2)$ which yields:
\begin{equation}\label{coeffNum}
\begin{split}
&\frac{M_{01} - r^\ast  M_{11}}{M_{20}}=\\
& = \left( -\frac{21\sqrt{\pi}}{\sqrt{2}} +
 3\left(- \gammast^{7/2}  +  (\gammast)^{3/2}\right)  K_{3/2}(\gammast) e^\gammast  +   
\left(  3(\gammast)^{5/2} -(\gammast)^{9/2} \right)  K_{5/2}(\gammast)  e^\gammast 
- (\gammast)^{7/2} K_{7/2}(\gammast)  e^\gammast \right)\\
&\times\left( 3\sqrt{2\pi} -  2 (\gammast)^{5/2}K_{5/2}(\gammast) e^\gammast  +  4 (\gammast)^{5/2}K_{3/2}(\gammast) e^{\gammast} \right)^{-1}\\
& \simeq 15.8363.
\end{split}
\end{equation}

\bibliography{bibRefsNew} 
\bibliographystyle{ieeetr}

\end{document}